\documentclass[twocolumn]{aastex631}
\usepackage{stfloats}
\usepackage{float}
\usepackage{graphicx}
\usepackage{natbib}
\usepackage{hyperref}
\usepackage{amsmath}
\usepackage{autobreak}
\usepackage{multirow}
\usepackage{makecell}
\usepackage{color}
\usepackage{bm}
\usepackage{threeparttable}
\usepackage[figuresright]{rotating}
\usepackage[mathscr]{euscript}
\usepackage[utf8]{inputenc}
\usepackage[T1]{fontenc}
\usepackage{amsmath}
\usepackage{makecell}
\usepackage{subfigure}
\usepackage{booktabs}

\begin{document}

\title{Exclusion of a direct progenitor detection for the Type~Ic SN~2017ein based on late-time observations}

\correspondingauthor{Ning-Chen Sun}
\email{sunnc@ucas.ac.cn}

\author{Yi-Han Zhao}
\affiliation{School of Astronomy and Space Science, University of Chinese Academy of Sciences, Beijing 100049, China}
\affiliation{National Astronomical Observatories, Chinese Academy of Sciences, Beijing 100101, China}

\author{Ning-Chen Sun}
\affiliation{School of Astronomy and Space Science, University of Chinese Academy of Sciences, Beijing 100049, China}
\affiliation{National Astronomical Observatories, Chinese Academy of Sciences, Beijing 100101, China}
\affiliation{Institute for Frontiers in Astronomy and Astrophysics, Beijing Normal University, Beijing, 102206, China}

\author{Junjie Wu}
\affiliation{School of Astronomy and Space Science, University of Chinese Academy of Sciences, Beijing 100049, China}
\affiliation{National Astronomical Observatories, Chinese Academy of Sciences, Beijing 100101, China}

\author{Zexi Niu}
\affiliation{School of Astronomy and Space Science, University of Chinese Academy of Sciences, Beijing 100049, China}
\affiliation{National Astronomical Observatories, Chinese Academy of Sciences, Beijing 100101, China}

\author{Xinyi Hong}
\affiliation{School of Astronomy and Space Science, University of Chinese Academy of Sciences, Beijing 100049, China}
\affiliation{National Astronomical Observatories, Chinese Academy of Sciences, Beijing 100101, China}

\author{Yinhan Huang}
\affiliation{School of Mathematical Sciences, University of Chinese Academy of Sciences, Beijing 100049, China}

\author{Justyn R. Maund}
\affiliation{Department of Physics, Royal Holloway, University of London, Egham, TW20 0EX, United Kingdom}

\author{Qiang Xi}
\affiliation{School of Astronomy and Space Science, University of Chinese Academy of Sciences, Beijing 100049, China}
\affiliation{National Astronomical Observatories, Chinese Academy of Sciences, Beijing 100101, China}

\author{Danfeng Xiang}
\affiliation{Beijing Planetarium, Beijing Academy of Sciences and Technology, Beijing 100044, China}

\author{Jifeng Liu}
\affiliation{School of Astronomy and Space Science, University of Chinese Academy of Sciences, Beijing 100049, China}
\affiliation{National Astronomical Observatories, Chinese Academy of Sciences, Beijing 100101, China}
\affiliation{Institute for Frontiers in Astronomy and Astrophysics, Beijing Normal University, Beijing, 102206, China}
\affiliation{New Cornerstone Science Laboratory, National Astronomical Observatories, Chinese Academy of Sciences, Beijing 100012, China}

\begin{abstract}

To date, SN\,2017ein is the only Type Ic supernova with a directly identified progenitor candidate. This candidate points to a very massive ($>$45~$M_\odot$) Wolf-Rayet progenitor, but its disappearance after the explosion of SN\,2017ein remains unconfirmed. In this work, we revisit SN\,2017ein in late-time images acquired by the Hubble Space Telescope (HST) at 2.4--3.8~yrs after peak brightness. We find this source has not disappeared and its brightness and color remain almost the same as in the pre-explosion images. Thus, we conclude that the pre-explosion source is not the genuine progenitor of SN\,2017ein. It is not much likely to be a companion star of the progenitor, since it has a much lower extinction than SN\,2017ein; its color also seems inconsistent with a star cluster, indicated by the newly added magnitude limit in F336W, apart from F555W and F814W. We suggest, therefore, this source is an unrelated star in chance alignment with SN\,2017ein. Based on the low ejecta mass, we propose that SN\,2017ein is most likely originated from a moderately massive star with $M_{\rm ini}$ $\sim$ 8--20 $M_\odot$, stripped by binary interaction, rather than a very massive Wolf-Rayet progenitor.
\end{abstract}

\defcitealias{kilpatrick2018potential}{K18}
\defcitealias{van2018sn}{VD18}
\defcitealias{Xiang2019ob}{X19}

\section{Introduction} \label{sec:intro}

Core-collapse supernovae (CCSNe) are violent explosive phenomena associated with massive stars at the end of their lives. Among them, Type~Ib SNe refer to those without H lines observed in their spectra, and Type~Ic SNe are those without H and He lines. It is believed that their progenitors have the outer envelopes stripped; some researchers also propose that Type~Ic SNe may still retain some He envelope, but the He lines are not apparent in the spectra due to a lack of excitation (\citealt{dessart2012nature}). In the local Universe, Type~Ib and Type~Ic SNe constitute $\sim$20\% of all CCSNe (\citealt{smith2011observed}).

Theoretically, Type~Ib/Ic SNe may arise from two possible progenitor channels (\citealt{woosley1997type}). In the single progenitor channel, a massive star with initial mass exceeding 25--30 $M_\odot$ can be stripped by its strong stellar winds (\citealt{conti1978mass}) and become a Wolf-Rayet (WR) star, characterized by extreme luminosity and prominent emission lines, before explosion.
In the binary progenitor scenario, a star may be stripped by a close companion via stable or unstable mass transfer involving Roche-lobe overflow and common-envelope evolution (\citealt{podsiadlowski2014evolution}). In the latter channel, the progenitors may be of lower initial masses compared with that of a single WR progenitor (\citealt{podsiadlowski1992presupernova}).

There have been direct progenitor detections in the pre-explosion images for $\sim$20 Type~II-P SNe (e.g.\newline SN\,2005cs, \citealt{maund2005progenitor}; SN\,2017eaw, \citealt{rui2019probing}; \citealt{van2019type}; SN\,2023ixf, \citealt{niu2023dusty}), 6 Type~IIn SNe (e.g. SN\,2005gl, \citealt{gal2007progenitor}; SN\,2010jl, \citealt{smith2011massive}; \citealt{fox2017candidate}; \citealt{niu2024disappearance}) and 6 Type~IIb SNe (e.g. SN\,2008ax, \citealt{crockett2008type}; \citealt{folatelli2015progenitor}; SN\,2017gkk, \citealt{niu2024discovery}). In contrast, the search for Type~Ib/Ic SN progenitors proves to be extremely difficult. Since their outer envelopes are stripped, the progenitors' spectral energy distributions (SEDs) peak at ultraviolet instead of optical wavelengths (\citealt{maund2018very}). In addition, Type~Ib/Ic SNe are often found in young star-forming regions with very high interstellar extinctions (\citealt{sun2023uv}). Up to now, only two progenitors of Type Ib SNe have been directly detected (i.e. iPTF13bvn, e.g. \citealt{cao2013discovery};  \citealt{bersten2014iptf13bvn}; SN\,2019yvr, \citealt{kilpatrick2023}; \citealt{sun2022environmental}), and only one progenitor candidate detection has been reported for a Type~Ic SN (i.e. SN\,2017ein, which shall be discussed later in this work; \citealt{kilpatrick2018potential} (\citetalias{kilpatrick2018potential}), \citealt{van2018sn} (\citetalias{van2018sn}), and \citealt{Xiang2019ob} (\citetalias{Xiang2019ob})).

The difficulty in direct progenitor detections motivates researchers to use alternative methods to determine the dominant evolutionary channel for the Type~Ib/Ic SN progenitors.
For example, \citet{smith2011observed} argued that the occurrence rate of Type~Ib/Ic SNe is too high to be compatible with the small number of massive WR stars at the high-mass end of the stellar initial mass function. 
\citet{lyman2016bolometric}, \citet{taddia2018carnegie} and \citet{woosley2021model} analyzed the light curves of stripped-envelope SNe (SESNe; Types IIb, Ib and Ic), all supporting interacting binaries as the dominant progenitor channel for Type~Ib/Ic SNe.
By analyzing the nebular-phase spectra and SN environments, \citet{fang2018supernova} and \citet{sun2023uv} suggested that Type~Ic SN progenitors may undergo a hybrid envelope-stripping mechanism, in which the hydrogen envelopes are stripped by binary interaction while the helium envelopes are stripped by intense stellar winds. \citet{solar2024binary}  confirmed observationally for a statistically large sample that Type~Ic SN progenitors are binary systems, using molecular clouds in the environments of the SN explosions as tracer of lifetimes.

SN\,2017ein is the only Type~Ic supernova with a progenitor detection. First discovered by Ron Arbour, it exploded in May 2017 in the nearby spiral galaxy NGC~3938, with a distance of $\sim$ 17 Mpc (\citealt{tully2009extragalactic}). \citetalias{kilpatrick2018potential}, \citetalias{van2018sn} and \citetalias{Xiang2019ob} conducted analyses of SN\,2017ein. They found its spectrum lacked hydrogen and helium lines, consistent with a Type~Ic classification; the peak magnitude in the V band is $-16\sim-17.5$ mag, dimmer than 80\% of Type~Ic SNe. These characteristics make it very similar to the carbon-rich, low-luminosity Type~Ic SN\,2007gr (\citealt{chen2014optical}). Although the Galactic extinction of SN\,2017ein is quite small of only $A_V$ = 0.058 mag (\citealt{schlafly2011measuring}), these studies found its host-galaxy extinction to be much larger, with estimated values ranging from $A_V$ = 0.94 to 1.84 mag, as measured with a variety of methods.

In the pre-explosion images of the Hubble Space Telescope (HST) , \citetalias{kilpatrick2018potential}, \citetalias{van2018sn}, and \citetalias{Xiang2019ob} found a point source located exactly at the SN position. They proposed that it is most likely to be the progenitor of SN\,2017ein. \citetalias{van2018sn} also suggested that it may be the binary companion or host star cluster of this SN. Any of these scenarios would suggest a very massive WR progenitor of > 45 $M_\odot$. The ejecta mass of SN\,2017ein was, however, only $\sim$ 1 $M_\odot$ (\citetalias{van2018sn} and \citetalias{Xiang2019ob}), which seems too low to be consistent with such a high progenitor mass (\citealt{lyman2016bolometric}, \citealt{taddia2018carnegie}). In addition, \citet{Jung2022} calculated the synthetic SEDs of massive WR progenitors with strong winds; they found the predicted colors are significantly different from that of the reported progenitor candidate of SN\,2017ein.

It is very important to confirm whether the reported candidate was the genuine progenitor by inspecting whether it has disappeared at late times \citep{2015MNRAS.447.3207M,niu2024disappearance}. In this work, we carry out such a study for the progenitor candidate of SN\,2017ein and re-examine its object type and relation with the SN. This paper is organized as follows. Section~\ref{sec:data} describes the data used in this work and how we conducted photometry. In Section~\ref{sec:results}, we present and discuss the results. This paper is then closed with a summary in Section~\ref{sec:sum}.

\section{Data and photometry}
\label{sec:data}

The data used in this work are from observations conducted with the HST (see Table~\ref{tab:data} for a complete list). The pre-explosion images were observed with the Wide Field Planetary Camera 2 (WFPC2) Planetary Camera (PC), while the late-time images were acquired with the Wild Field Camera 3 (WFC3) Ultraviolet-Visible (UVIS) channel at t = 2.4--3.8 years after the peak brightness. All the observations were conducted in the F555W and F814W filters, with an addition of F336W at t = 2.5 yrs. Besides, an F438W image, taken right after the SN explosion, was used as a reference image for differential astronomy. The HST data presented in this article were obtained from the Mikulski Archive for Space Telescopes (MAST) at the Space Telescope Science Institute. The specific observations analyzed can be accessed via \dataset[https://doi.org/10.17909/7n8j-cy13]{https://doi.org/10.17909/7n8j-cy13}.

We retrieved the data from the Barbara A. Mikulski Archive for Space Telescopes (MAST; \url{https://mast.stsci.edu}). For the pre-explosion WFPC2 images, we used the \textsc{iraf} task \textsc{crrej} (\citealt{gonzaga_wfpc2_3.0}) to remove cosmic rays. For the late-time WFC3 images, we used the \textsc{tweakreg} (\citealt{avila2014drizzlepac}) package to align the individual exposures (\texttt{*\_flc.fits} images) and combined them with the \textsc{astrodizzle} (\citealt{hack2012astrodrizzle}) package to remove the cosmic rays.

The \textsc{dolphot} (\citealt{dolphin2000wfpc2}) package was used to conduct point-source photometry. In order to see how the photometry parameters may influence the final results, we tested the use of different parameters of \texttt{FitSky = 2} or \texttt{3}, \texttt{RAper = 3} or \texttt{8}, \texttt{ApCor = 0} or \texttt{1}, \texttt{Force1 = 0} or \texttt{1}. We find that the results do not change significantly with the chosen values. In this work, we adopted the result obtained with \texttt{FitSky = 2}, \texttt{RAper = 3}, \texttt{ApCor = 1}, \texttt{Force1 = 1}, as recommended for crowded fields. We also used common stars to correct for the possible systematic errors between observations at different epochs; the details are provided in Appendix \ref{sys.sec}.

\begin{table*}[!htbp]
\centering
\caption{HST data and photometry. \label{tab:data}}
    \begin{tabular}{ccccccccc}
    \hline
    \hline
        \toprule
         Date (UT) & Time from & Instrument & Filter & Exposure & Program & Systematic error in & Magnitude$^c$ \\
          & peak (yr) & & & Time (s) & ID & photometry$^{b}$ (mag) & (mag) \\
        \midrule
        2007 Dec 12.4 & $-10$ & WFPC2/PC & F555W & 460 & 10877 & $-$0.37 (0.01) & 24.75 (0.11) \\
        2007 Dec 12.4 & $-10$ & WFPC2/PC & F814W & 700 & 10877 & $-$0.26 (0.01) & 24.56 (0.18)\\
        2017 Jun 12.1 & $0$ & WFC3/UVIS & F438W & 270 & 14645 & -- & --$^{d}$ \\ 
        2019 Nov 7.7 & $2.4$ & WFC3/UVIS & F555W & 1913$^{a}$ & 15691 & $-$0.03 (0.004) & 24.83 (0.05)\\
        2019 Nov 7.8 & $2.4$ & WFC3/UVIS & F814W & 1272$^{a}$ & 15691 & $-$0.08 (0.006) & 24.58 (0.14)\\
        2019 Dec 1.0 & $2.5$ & WFC3/UVIS & F336W & 2625 & 15853 & -- & 23.61 (0.06)\\
        2019 Dec 7.1 & $2.5$ & WFC3/UVIS & F555W & 1913$^{a}$ & 15691 & $-$0.05 (0.004) & 24.77 (0.06)\\
        2019 Dec 7.2 & $2.5$ & WFC3/UVIS & F814W & 2544 & 15691 & $-$0.06 (0.005) & 24.50 (0.05)\\
        2021 Apr 2.9 & $3.8$ & WFC3/UVIS & F555W & 710 & 16239 & -- & 24.72 (0.07)\\
        2021 Apr 2.9 & $3.8$ & WFC3/UVIS & F814W & 780 & 16239 & -- & 24.76 (0.14)\\
        \bottomrule
    \end{tabular} \\
    $a$ Some exposures are discarded due to bad image quality and are not included in the exposure time as listed here.\\
    $b$ The values are relative to observations on $t = 3.8$ yrs. \\
    $c$ The systematic errors have been corrected except for F336W.\\
    $d$ This image is only used as a reference image for differential astrometry.
\end{table*}

\section{Results and discussion}
\label{sec:results}

\begin{figure*}[!htbp]
    \centering
    \includegraphics[width=0.80\textwidth]{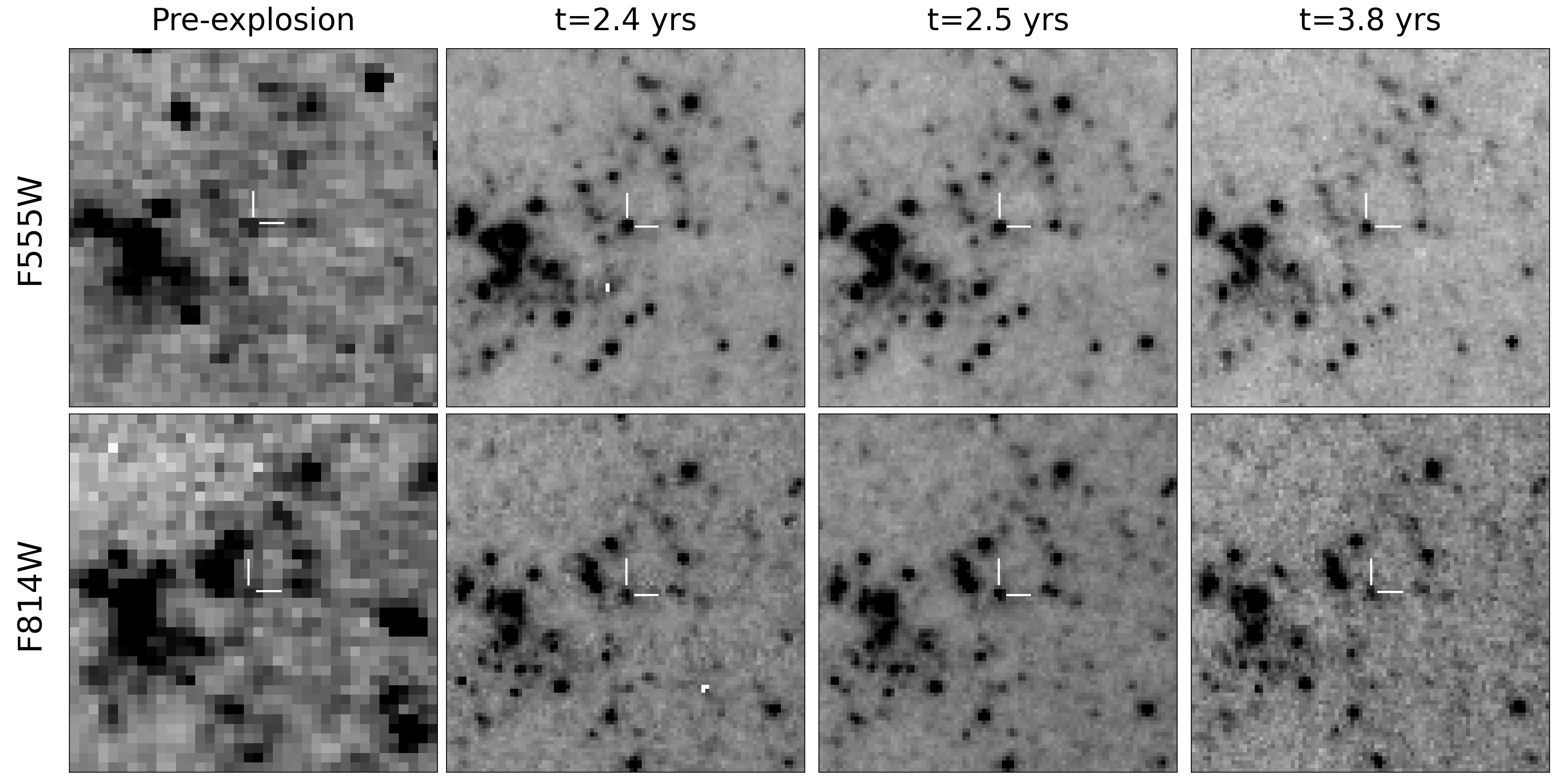}
    \caption{\enspace Pre- and post-explosion HST images of the site of SN\,2017ein (F336W is not included here). The SN position is denoted by the white crosshair. Each image has a size of $\sim$ 160$\times$160 pc and is oriented with North up and East to the left.}
    \label{fig:images}
\end{figure*}

Figure~\ref{fig:images} shows the site of SN\,2017ein in the pre-explosion and late-time observations. For better determination of the SN position, differential astrometry was used to align the images with respect to the image taken near peak brightness. We selected more than ten common stars as reference stars, and the uncertainty of the alignment reaches 0.7 pixel. It is clearly shown that in the pre-explosion images, there is a source at the SN position in both F555W and F814W bands, with a positional difference of only $\sim$0.13~pixel. This source corresponds to the progenitor candidate of SN\,2017ein as reported in the previous studies (\citetalias{kilpatrick2018potential}, \citetalias{van2018sn}, and \citetalias{Xiang2019ob}). For convenience, we shall refer to it as \textit{Source~A} hereafter. In the late-time images, however, \textit{Source~A} does not disappear and the SN position remains still bright. The sharpness parameter of \textit{Source~A} reported by \textsc{dolphot} is very close to zero, which implies that \textit{Source~A} is a point source.

\begin{figure}[h]
    \centering
    \includegraphics[width=1\linewidth]{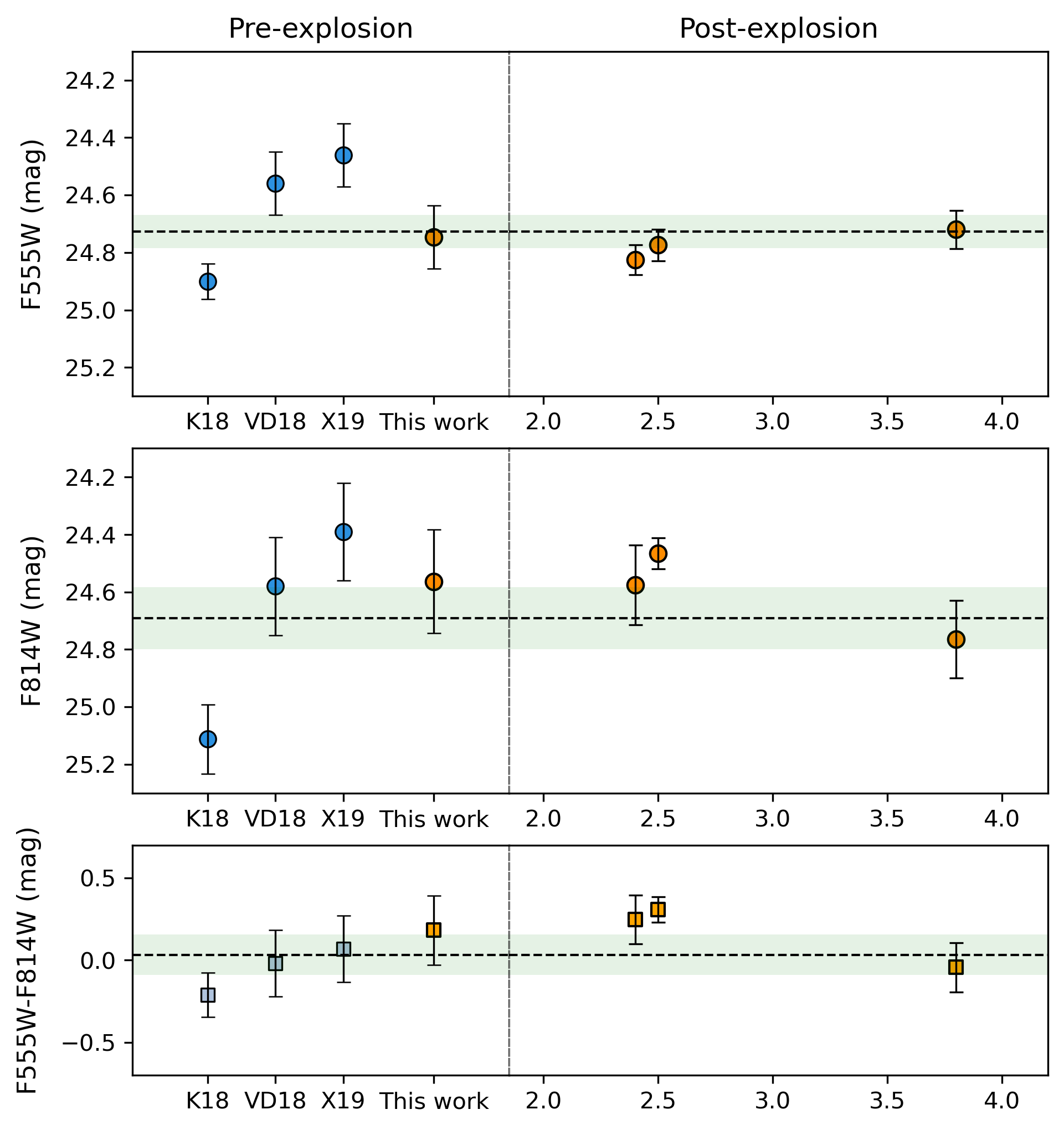}
    \caption{\enspace Brightness and color evolution of \textit{Source~A} at different epochs. The pre-explosion magnitudes measured by  \citetalias{kilpatrick2018potential}, \citetalias{Xiang2019ob} and \citetalias{van2018sn} are also plotted for comparison. The dashed lines represent the inverse-variance weighted averages of the pre-explosion and the last epoch, and the green-shaded bands indicate the errors. All the magnitudes have been corrected for systematic bias between the different epochs.}
    \label{fig:Light curves}
\end{figure}

In Figure~\ref{fig:Light curves}, we show the brightness and color evolution of \textit{Source~A} . Note that the possible systematic errors between different epochs have been corrected. The pre-explosion magnitudes reported by \citetalias{kilpatrick2018potential}, \citetalias{van2018sn}, and \citetalias{Xiang2019ob} are also included for comparison. It is worth noting that the values reported by different groups vary slightly from each other, which may arise from different photometry methods. We derive pre-explosion magnitudes of F555W = 24.75 $\pm$ 0.11 mag, F814W = 24.56 $\pm$ 0.18 mag, which are generally consistent with the previous results except for the F814W magnitude reported by \citetalias{kilpatrick2018potential}. The late-time F555W magnitudes at $t$ = 2.4, 2.5, 3.8~yrs remain almost the same as the pre-explosion level. In the F814W images, the late-time brightness is also consistent with the pre-explosion level except for that at $t$ = 2.5~yrs, which has a high signal-to-noise ratio and exceeds by $\sim$2$\sigma$. The pre-explosion color and that at the last epoch are consistent with F555W $-$ F814W $\sim$ 0~mag within uncertainties; in contrast, the color at $t$ = 2.5~yrs appears slightly redder than 0 by more than 1$\sigma$.

According to these phenomena, the most likely scenario is that the pre-explosion object, a star or star cluster, has not disappeared after the explosion of SN\,2017ein since the late-time brightness and color are very consistent with the pre-explosion ones. The extra light in F814W and the red color at $t$ = 2.5~yrs could arise from a light echo (\citealt{Boffi1999}; \citealt{bond2003energetic}) or CSM interaction (\citealt{Pellegrino_2022}; \citealt{Fraser2020}), which has completely faded by the last epoch. A light echo or CSM interaction could also be present at $t$ = 2.4~yrs, but the observations have such large photometric uncertainties that we could not identify their contribution. It is noteworthy that another scenario is also considered, where the SN ejecta shocks the surviving binary companion, causing it to be over luminous and inflated for several years after the SN explosion \citep{hirai2018comprehensive,ogata2021,chen2023}. However, in such a scenario, there is little possibility that \textit{Source~A}'s magnitudes of different bands remain this consistent with the pre-explosion ones.

\begin{figure}
    \centering
    \includegraphics[width=0.9\linewidth]{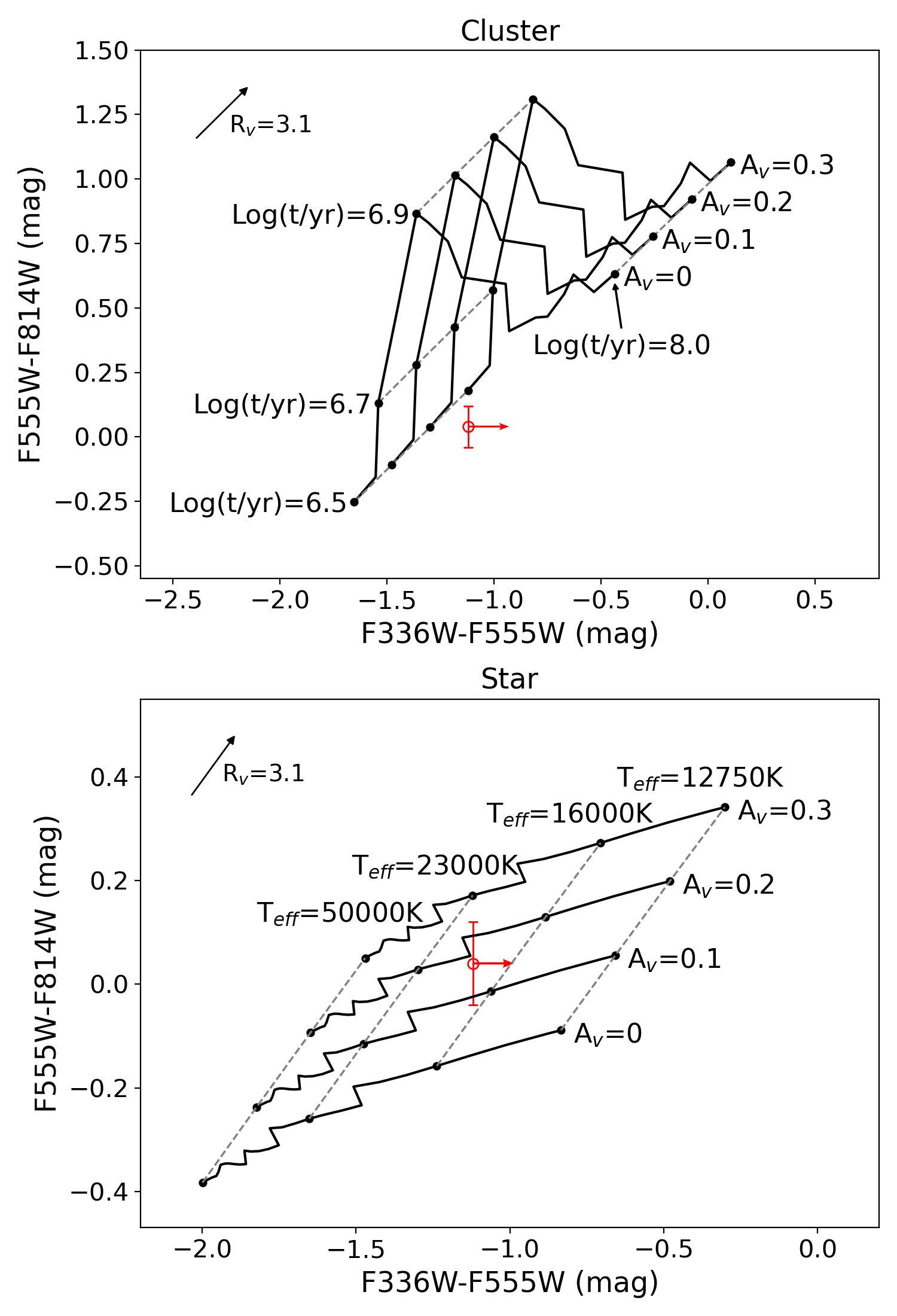}
    \caption{\enspace F336W$-$F555W v.s. F555W$-$F814W color-color diagrams for \textit{Source~A} (red circle) and model stars (bottom panel, solid lines) and star clusters (top panel, solid lines). The arrows in the upper-left corner show the reddening vector under a standard extinction law with $R_V$ = 3.1.}
    \label{fig:cc diagram}
\end{figure}

We analyze \textit{Source~A} on the (F336W$-$F555W, F555W$-$F814W) color-color diagram (Figure~\ref{fig:cc diagram}) to determine whether it is a star or a star cluster. To reduce the photometric uncertainties in F555W/F814W, we use the inverse-variance weighted averages of the pre-explosion magnitude and that at the last epoch as the adopted magnitude of \textit{Source~A} (i.e. 24.73 $\pm$ 0.06 mag and 24.69 $\pm$ 0.11 mag, respectively). For F336W, we conservatively regard the derived magnitude of 23.61 $\pm$ 0.06 mag as a brightness upper limit for the underlying star or star cluster, since this magnitude is measured at t = 2.5 yrs, by which time there could still be some extra light caused by a light echo or CSM interaction. This possible extra light may also influence the brightness in F555W and F814W, but the effects are believed to be negligible (see above discussion). For comparison, we show synthetic colors for stars of different effective temperatures (based on the \citealt{castelli2004new} stellar atmospheric models) and star clusters of different ages (based on the \textsc{bpass} model population synthesis; \citealt{eldridge2017binary}; \citealt{stanway2018re}; \citealt{byrne2022dependence}). For both stellar and cluster SEDs, we also apply a range of different extinctions. 

In Fig.~\ref{fig:cc diagram}(a), the color of \textit{Source~A} appears very different from that expected for star clusters. The previous studies (\citetalias{kilpatrick2018potential}, \citetalias{van2018sn}, and \citetalias{Xiang2019ob}) also discussed the possibility of \textit{Source~A} being a young star cluster but only with the F555W $-$ F814W color from the pre-explosion observations. Now with the addition of the F336W limit, we suggest this scenario is not very likely since the young star clusters would be too bright in this filter. This analysis, however, is based on a standard extinction law of $R_V$ = 3.1; the color of \textit{Source~A} could still be consistent with that of a star cluster in case of a different host-galaxy extinction law. Moreover, since \textit{Source~A} lies quite close to the cluster models on the diagram, we might also consider the possibility of it truly being a cluster, with an age of $10^{6.5}
$ yrs and a total extinction of $A_V = 0.3$~mag. We estimated the mass of such a cluster to be $10^3$ $M_\odot$. Binaries or single stars with $M < 114 M_\odot$ can explode at this age, based on BPASS models. It is noteworthy that this extinction value is much lower than that of SN\,2017ein (1.26 $\pm$ 0.2 mag, \citetalias{kilpatrick2018potential}; 1.0--1.9~mag, \citetalias{van2018sn}; 1.29 $\pm$ 0.18 mag, \citetalias{Xiang2019ob}; measured with different methods), which might be explained by the supposition that the SN explodes from a dusty star within the cluster. Moreover, we find the influence of metallicity on the cluster models is negligible in our work.

In Fig.~\ref{fig:cc diagram}(b), the color of \textit{Source~A} appears consistent with those of stellar models with effective temperature of $T_{\rm eff} \lesssim$ 16000--21000~K and (total) extinction of $A_V < 0.23$~mag, which suggests that \textit{Source~A} is most likely to be a hot massive star. However, it is not the progenitor star of SN\,2017ein since it has not vanished after explosion. It is worth noting that the derived total extinction of \textit{Source~A} is much smaller than that of SN\,2017ein itself ($A_V > 1$, as mentioned before). We argue therefore that it is not a companion star of SN\,2017ein but an unrelated star in chance alignment. We calculated the probability of such a chance alignment as 4.8$\%$, according to the spatial density of sources seen in Fig.~\ref{fig:images}.

\citet{Jung2022} noticed that \textit{Source~A} is too blue and too bright to align with any of the stellar atmospheric models that they developed for single WR progenitors of Type~Ic SNe. This can now be explained by our work, as it turns out that \textit{Source~A} is not the genuine progenitor but most likely an unrelated star in chance alignment.

\citetalias{kilpatrick2018potential}, \citetalias{van2018sn}, and \citetalias{Xiang2019ob} suggested that \textit{Source~A} could be the progenitor of SN\,2017ein, or its companion star, or a host star cluster. In any scenario, \textit{Source~A} would support a very massive WR progenitor with an initial mass of $>$45 $M_\odot$. In this work, however, we find \textit{Source~A} is actually an unrelated star in chance alignment. Based on the low ejecta mass ($\sim$1 $M_\odot$; \citetalias{van2018sn}; \citetalias{Xiang2019ob}), we suggest the progenitor of SN\,2017ein is most likely to be a moderately massive star of 8--20~$M_\odot$ stripped in a binary system.

\section{Summary}\label{sec:sum}

In this work, we re-investigate the progenitor candidate reported for SN\,2017ein with the new HST images acquired at 2.4--3.8~yrs after peak brightness. We find that the candidate has not disappeared, with its brightness and color almost the same as the pre-explosion ones. Thus, we conclude that the candidate is not the genuine progenitor. It is neither a companion star of the progenitor considering its much lower extinction than SN\,2017ein. The color-color diagram analysis also indicates that this source is less likely to be a star cluster. Therefore, we suggest it is most likely to be an unrelated star in chance alignment with SN\,2017ein. Based on the low ejecta mass, we suggest that the progenitor is more likely to be a moderately massive star stripped in a binary, rather than a very massive WR star as suggested by previous works.

\section*{acknowledgments}
This work is supported by the Strategic Priority Research Program of the Chinese Academy of Sciences, Grant No. XDB0550300. NCS’s research is funded by the NSFC grants No. 12303051 and No. 12261141690 and ZXN acknowledges support from the NSFC through grant No. 12303039. JFL acknowledges support from the NSFC through grants No. 11988101 and No. 11933004 and from the New Cornerstone Science Foundation through the New Cornerstone Investigator Program and the XPLORER PRIZE. We thank the anonymous referee for the helpful comments.

\appendix
\section{Systematic error correction}
\label{sys.sec}

To correct for possible systematic biases between different epochs, we selected common stars and compared their photometry. The observations at the last epoch of $t$ = 3.8~yrs were used as reference.
Common stars were selected with a minimum signal-to-noise ratio of 5 for WFPC2 and 10 for WFC3 observations; and since WFPC2 has a lower spatial resolution, we required the stars to be isolated without any neighboring sources within 0.24~arcsec. With these two criteria, hundreds of stars were selected for comparison.
The results are shown in Figure~\ref{fig:syserr} and Figure~\ref{fig:Hsyserr}. We find that there are obvious systematic errors in the pre-explosion WFPC2 observations, while the systematic errors at other epochs of WFC3 observations are much smaller and negligible. We calculated the inverse-variance weighted averages with 3$\sigma$ clipping as an estimate of the systematic errors. The values for each epoch are listed in Table~\ref{tab:data}.

\setcounter{figure}{0}
\renewcommand{\thefigure}{A\arabic{figure}}
\begin{figure*}[!htbp] 
    \centering
    \includegraphics[width=1.0\textwidth]{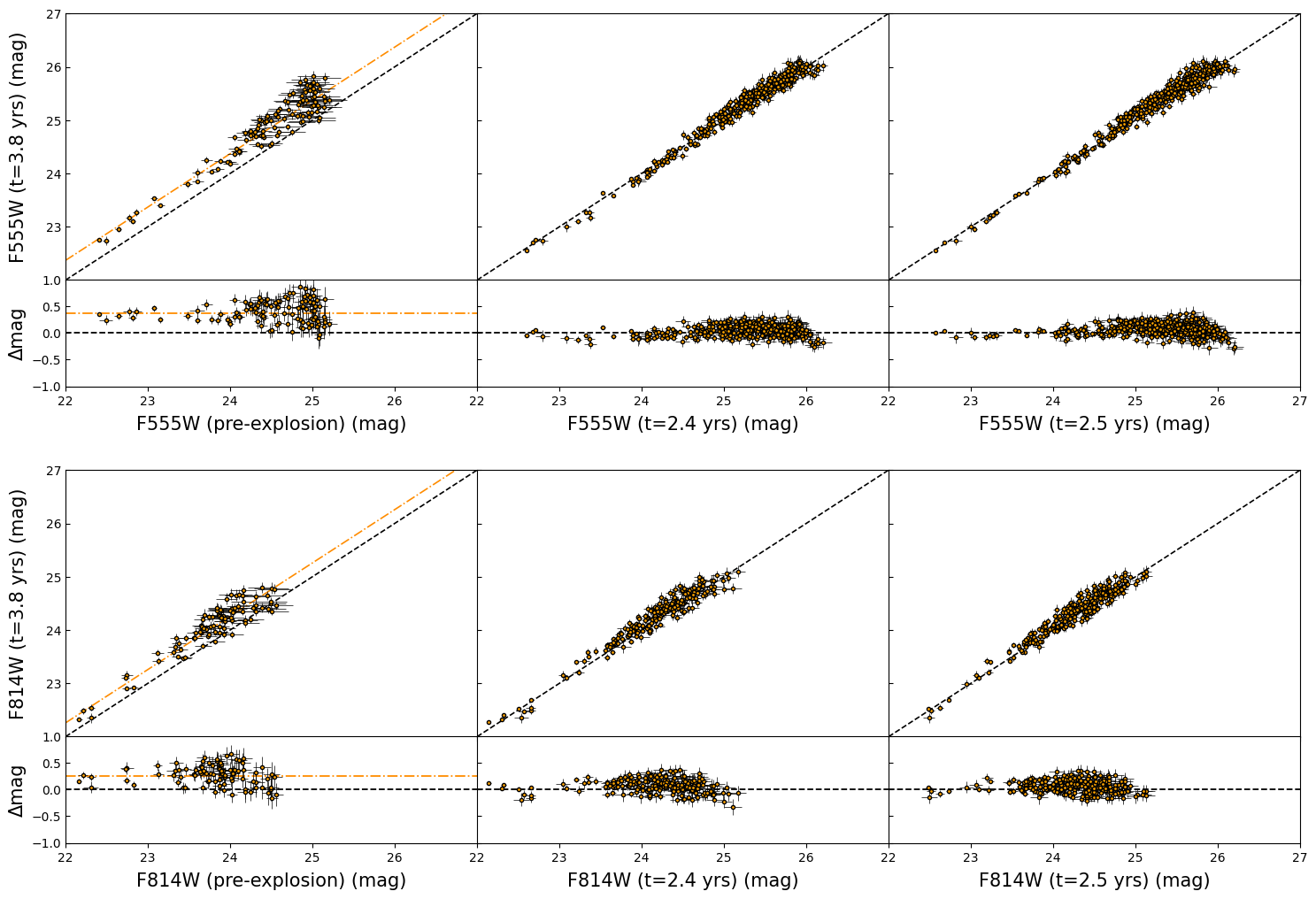}
    \caption{\enspace Comparison of photometry as reported by \textsc{dolphot} for observations at different epochs. The black dashed line shows the one-to-one relation while the orange dash-dotted line corresponds to systematic shifts as calculated by inverse-variance weighted averages.}
    \label{fig:syserr}
\end{figure*}

\setcounter{figure}{1}
\renewcommand{\thefigure}{A\arabic{figure}}
\begin{figure*}[!htbp] 
    \centering
    \includegraphics[width=1.0\textwidth]{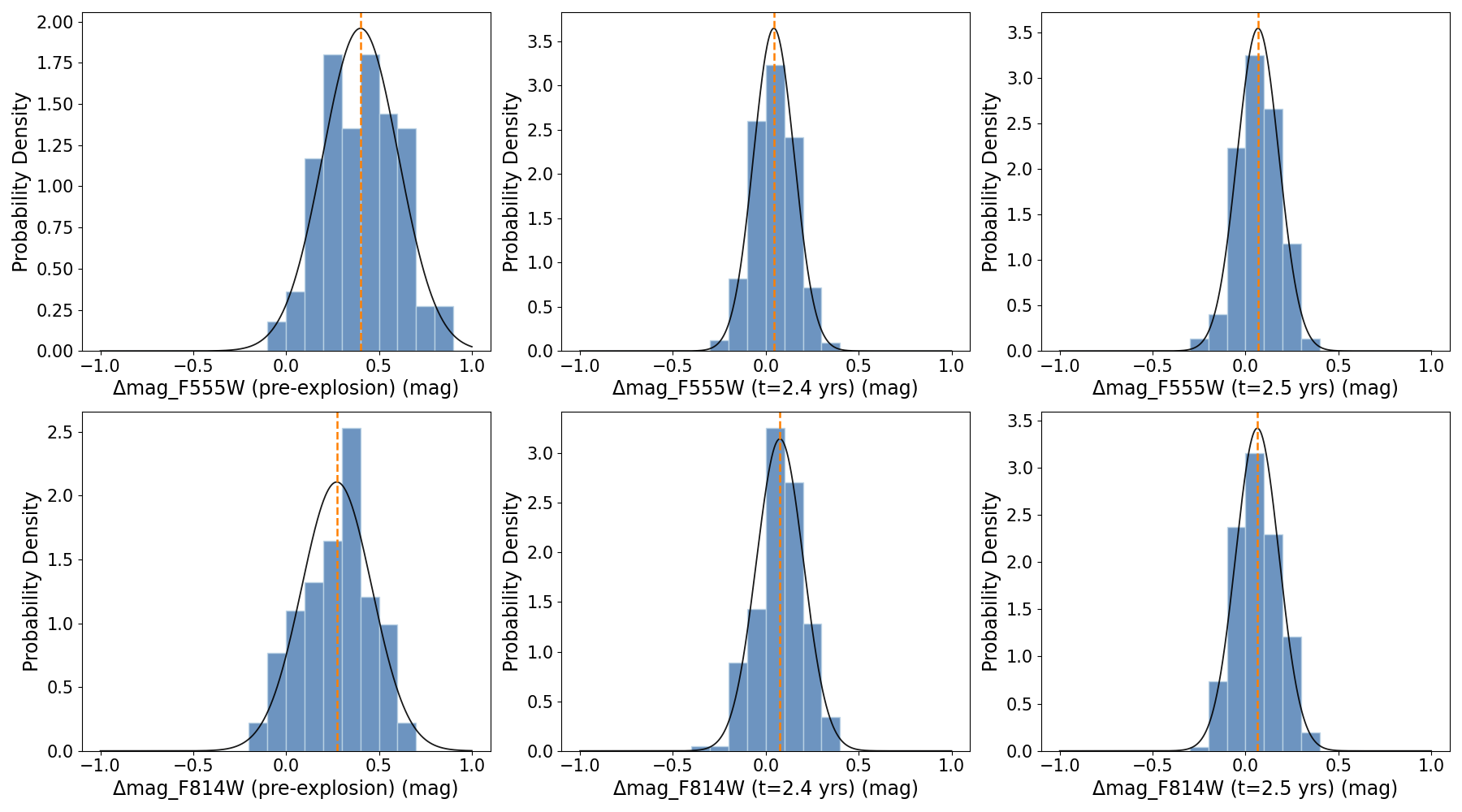}
    \caption{\enspace Histograms of magnitude differences with respect to the epoch of t = 3.8 yrs. The solid lines are Gaussian fit curves and the dashed lines represent the means of the Gaussian fit. Normalization has been applied to each graph.}
    \label{fig:Hsyserr}
\end{figure*}

\vspace{5mm}

\bibliography{SN2017ein}{}
\bibliographystyle{aasjournal}

\end{document}